\documentclass[]{aa}
\usepackage[utf8]{inputenc}
\usepackage{graphicx}
\usepackage{txfonts}
\begin{document}

   \title{New constraints on the values of the fundamental constants at a look-back time of 7.3 Gyr}

   \author{Renzhi Su \inst{1,2}, 
          Stephen J., Curran\inst{3}, 
          Francoise Combes\inst{4}, 
          Neeraj Gupta\inst{5}, 
          Sebastien Muller\inst{6}, 
          Di Li\inst{7,8,2} 
          \and
          Minfeng Gu\inst{1}          
          }

   \institute{Shanghai Astronomical Observatory, Chinese Academy of Sciences, 80 Nandan Road, Shanghai 200030, China\\
              \email{rzsu.astro@gmail.com}
              \and
            Research Center for Astronomical Computing, Zhejiang Laboratory, Hangzhou 311100, China
         \and
             School of Chemical and Physical Sciences, Victoria University of Wellington, PO Box 600, Wellington 6140, New Zealand
         \and
             Observatoire de Paris, LERMA, Coll\`ege de France, CNRS, PSL University, Sorbonne University, 75014, Paris, France
         \and
             Inter-University Centre for Astronomy and Astrophysics, Post Bag 4, Ganeshkhind, Pune, 411 007, India
         \and
             Department of Space, Earth and Environment, Chalmers University of Technology, Onsala Space Observatory,
             43992 Onsala, Sweden
         \and
             New Cornerstone Science Laboratory, Department of Astronomy, Tsinghua University, Beijing 100084, China
        \and
             National Astronomical Observatories, Chinese Academy of Sciences, Beijing 100012, China
             }

   \date{Received XX XX, 202X; accepted XX XX, 202X}

% \abstract{}{}{}{}{} 
% 5 {} token are mandatory
 
\abstract
  % context heading (optional)
  % {} leave it empty if necessary  
   {The study of redshifted spectral lines can provide a measure of the fundamental constants over large look-back times. Current grand unified theories predict an evolution in these constants and astronomical observations offer the only experimental measure of the values of the constants over large timescales. Of particular interest are the dimensionless constants: the fine structure constant ($\alpha$), the proton-electron mass ratio ($\mu$), and the proton g-factor ($g_p$), since these do not require a "standard meterstick". Here we present a re-analysis of the 18 cm hydroxyl (OH) lines at $z=0.89$, which were recently detected with the MeerKAT telescope, toward the radio source PKS\,1830-211. Utilizing the previous constraint of $\Delta\mu/\mu=(-1.8\pm1.2)\times10^{-7}$, we obtain $\Delta (\alpha g_p^{0.27})/(\alpha g_p^{0.27})\lesssim5.7\times10^{-5}$, $\Delta \alpha/\alpha\lesssim2.3\times10^{-3}$ , and $\Delta g_p/g_p\lesssim7.9\times10^{-3}$. These new constraints are consistent with no evolution over a look-back time of 7.3 Gyr and provide another valuable data point in the putative evolution of the constants. 
}

   \titlerunning{New constraints on fundamental constants toward PKS 1830-211}
   \authorrunning{Renzhi Su et al.}
   \maketitle
%
%-------------------------------------------------------------------

\section{Introduction} \label{sec:intro}
Cosmological variations in fundamental constants
are predicted in higher-dimensional theories that aim to unify the standard model of particle physics and general relativity \citep[e.g.,][]{marciano1984,damour1994,li1998}. 
Terrestrial measurements of fractional changes in these constants that employ the Oklo natural fission reactor and 
atomic clocks have been used to measure the fine structure constant, $\alpha$ \citep[e.g.,][]{damour1996,rosenband2008}, but they are limited to timescales of two billion and only a few years, respectively. 

Through the finite speed of light, astronomical observations offer the possibility of measuring the values of the fundamental constants over much larger look-back times. Furthermore, these can also constrain models of cosmology \citep[e.g.,][]{sandvik2002,thompson2013,thompson2013_2}, in addition to constructing the equation of state of dark energy \citep[e.g.,][]{avelino2006,nunes2009}, thereby transforming our basic understanding of physics and the Universe. Studies of the cosmological evolution of fundamental constants are thus of great interest \citep[see the reviews by][]{uzan2003,uzan2011,uzan2024}. 

The measurements of the values of the fundamental constants are obtained by comparing the measured redshifts between various spectral transitions with rest frequencies, each of which has its own unique dependence on the combination of fundamental constants \citep[e.g.,][]{chengalur2003,darling2003,curran2004}. Much of the previous work has focused on optical and ultra-violet spectral lines toward quasi-stellar objects (QSOs), but the results have been inconclusive \citep[e.g.,][]{murphy2004,murphy2003,murphy2003_2,srianand2004,murphy2008_2,griest2010}. However, interpretation of the data is affected by
relatively coarse spectral resolution and wavelength calibration, of which radio/millimeter band observations are largely free. Therefore, observations in these bands can provide more reliable measurements
\citep[e.g.,][]{darling2004,flambaum2007,murphy2008,kanekar2011,curran2011,kanekar2012,bagdonaite2013,kanekar2018,su2025}.

Like the optical--UV studies, the comparisons of different species, such as neutral hydrogen ($\mbox{H\,{\sc i}}$), OH, millimeter lines, and optical--UV lines, are subject to physical velocity offsets between species, which can cause an offset in redshifts \cite[e.g.,][]{tzanavaris2005,tzanavaris2007}. Thus, with different transitions, each of which has a different dependence on the fundamental constants, OH is of particular interest since the spectra are obtained from the same gas and they can simultaneously measure the variations of $\alpha$, $\mu$, and $g_{p}$ \cite[e.g.,][]{darling2004,kanekar2004}. However, the detection of redshifted OH absorption remains scarce, being limited to five at $z>0.1$, with all at $z<0.89$
\citep[e.g.,][and references therein]{curran2007,gupta2018,zheng2020,curran2021,chandola2024}. The highest redshift example is in the lensing galaxy at $z=0.8858$ toward the $z=2.507$ quasar PKS\,1830--211 \citep{wiklind1996}. In this absorber, atomic (\mbox{H\,{\sc i}}) and
various molecular (e.g., OH, $\rm CH_{3}OH$) species have been detected \citep[e.g.,][]{chengalur1999,koopmans2005,ellingsen2012,bagdonaite2013,kanekar2015,schulz2015,allison2017,tercero2020,gupta2021,combes2021,muller2011,muller2014,muller2020,muller2023,muller2024}. 

Comparing different transitions from methanol ($\rm CH_{3}OH$) in PKS\,1830--211 has given the stringent constraint $\Delta\mu/\mu=(-1.8\pm1.2)\times10^{-7}$ \citep{muller2021}. Using this in conjunction with recent MeerKAT observations of the optical-thin thermalized OH 18 cm lines and the optical-thin conjugate OH 18 cm satellite lines \citep{combes2021}\footnote{In the local thermodynamic equilibrium, the relative strengths of OH 18 cm lines are 1612:1665:1667:1720 MHz = 1:5:9:1 and, where the pumping is dominated by the intraladder 119~$\mu$m transition, the OH 18 cm satellite lines will appear as stimulated absorption (1612 MHz) and emission (1720 MHz) with the same strength, showing conjugate behavior \cite[e.g.,][]{darling2004,kanekar2018,combes2021}. Under the latter condition, the OH 18 cm main lines are suppressed, exhibiting no absorption or emission.}, 
we add new constraints to the cosmological evolution of $\alpha$ and $g_p$.
Throughout this paper, we use the $\Lambda$CDM cosmology with $H_{0}=70\,$km$\,$s$^{-1}\,$Mpc$^{-1}$,  $\Omega_{m}$=0.3, and $\Omega_{\Lambda}$=0.7.

\section{The data} \label{sec:data}

The data were obtained from \cite{combes2021}. MeerKAT wideband L- and UHF-band observations  were performed toward  PKS\,1830--211 in 2019 \citep{gupta2021} and 2020  \citep{combes2021}, respectively, as part of the MeerKAT Absorption Line Survey \citep[MALS;][]{gupta2016}. The L-band observations clearly detected the OH 18 cm main lines and tentatively detected the OH 1720 MHz line, and the UHF-band observations detected all four lines. The OH 1612, 1665, and 1667~MHz spectra presented in \cite{combes2021} are from the UHF-band data and the 1720~MHz line spectrum is from a combination of the L- and UHF-bands, weighted based on their root-mean-square (rms) values. They have a velocity resolution of 5.8, 5.6, 5.6, and 5.4 $\rm km~s^{-1}$ with a normalized rms noise of 0.00027, 0.00028, 0.00028, and 0.00020, respectively.

\section{Line fitting} \label{sec:line_fit}

The line strength ratio between the 1665 and 1667~MHz (main) lines is consistent with 5:9, which indicates thermalized OH gas, although the 1612~MHz absorption is stronger than expected from local thermodynamic equilibrium (LTE). This is accounted for by conjugate OH satellite lines, which also explains the observed OH 1720 MHz emission \citep{combes2021}.

\citeauthor{combes2021} jointly fitted three Gaussian profiles to the OH 18 cm main lines. However, one of these is very weak. Furthermore, the possible frequency evolution of the lines was not considered. If we consider the evolution of the fundamental constants, the frequency discrepancy between observed lines is different from that derived from laboratory frequencies. We therefore re-fitted the data according to the following criteria:

- We used two independent Gaussian functions to fit the OH 1667 MHz line. Under LTE, we can determine the corresponding counterparts in other OH 18 cm lines. The strength and full width at half maximum (FWHM) of these Gaussian functions were tightened.
    
- To account for the possible frequency evolution between the OH 1665 MHz and OH 1667 MHz lines, we added a frequency offset, FO1, to the frequency of the expected OH 1665 MHz line. Similarly, we added another velocity offset, FO2, to the frequency of the expected OH 1612 MHz line to account for the possible frequency evolution between the OH 1612 MHz and OH 1667 MHz lines. Since, in principle, the frequencies of four OH 18 cm lines have the relation $\nu_{1612}+\nu_{1720}=\nu_{1665}+\nu_{1667}$, with FO1 and FO2, we can determine the frequency evolution between the OH 1720 MHz and OH 1667 MHz lines.

- To account for the extra absorption in the OH 1612 MHz line contributed by the conjugate OH 18 cm satellite lines, we used two additional independent Gaussian functions to fit the OH 1612 MHz line. As expected from conjugate behavior, we needed to fit two other Gaussian functions with reverse intensity to the OH 1720 MHz line.

- The four OH 18 cm lines were fitted simultaneously.\newline

\begin{figure*}
        \includegraphics[width=\textwidth]{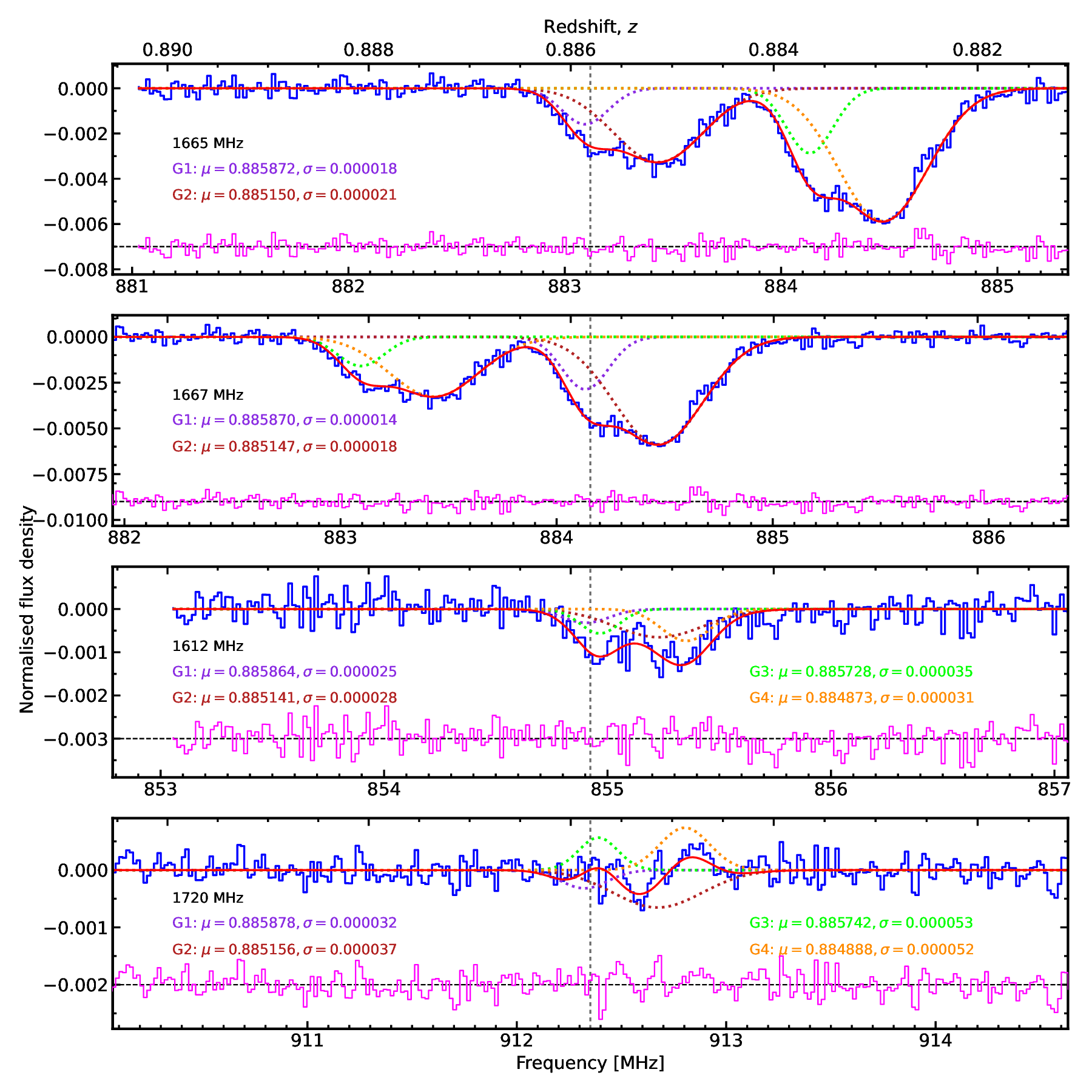}
    \caption{Gaussian fitting of the OH 18 cm lines. From top to bottom are the OH 1665, 1667, 1612, and 1720 MHz lines. Two Gaussian functions fitted to the OH 1667 MHz line are marked with dotted purple and magenta lines. The corresponding counterparts of these two Gaussian components in other lines in the LTE condition are also marked with dotted purple and magenta lines. The conjugate OH satellite lines (absorption in the 1612 MHz line and emission in the 1720 MHz line) are marked with green and orange. The vertical dashed gray line marks the optical redshift. In all subplots, violet lines represent residuals that have been shifted in the Y axis for clarity; the redshifts and their associated uncertainties of individual components are denoted with $\mu$ and $\sigma$.}
    \label{fig:spec_fit}
\end{figure*}

Hence we had  14 free parameters in the fitting. Four Gaussian functions were fitted to the OH 1612 MHz line, two were fitted to the OH 1665 MHz line, two were fitted to the OH 1667 MHz line, and four were fitted to the OH 1720 MHz line. The fitting gave a reduced $\chi^{2}$ of 1.17 and the residuals are consistent with noise. We tried to fit the OH 1667 MHz line with three Gaussian functions, but the fitting could not be improved. The spectra fitting results are shown in Fig. \ref{fig:spec_fit} and the derived parameters for the OH 18 cm lines are listed in Table \ref{tab:fit_results}, where the frequency offsets F01 and F02 have been added into the center frequencies of OH 1612, 1665, and 1720 MHz lines.

\begin{table*}
\caption{Derived parameters for the OH 18 cm lines. }
\centering
\label{tab:fit_results}
\begin{tabular}{lclcc}
 \hline
Line& Name of OH gas  & Normalized Amplitude & Center & FWHM \\
&  & & (MHz)&($\rm km\,s^{-1}$)  \\
 \hline
OH 1612   & G1& -0.00287(23)$\times$1/9 & 854.9033(115) & 36.5$\pm$2.3 \\
          & G2& -0.005880(71)$\times$1/9 & 855.2310(126) &69.7$\pm$2.2  \\
          & G3&-0.000566(74)  & 854.9650(157) & 30.7$\pm$5.3 \\
          & G4&-0.000739(68)  &855.3526(141)  & 39.0$\pm$4.5 \\

\hline
OH 1665   & G1& -0.00287(23)$\times$5/9 & 883.09362(838) & 36.5$\pm$2.3 \\
          & G2& -0.005880(71)$\times$5/9 & 883.43211(995) & 69.7$\pm$2.2 \\
\hline
OH 1667   & G1& -0.00287(23) & 884.13264(640) & 36.5$\pm$2.3 \\
          & G2& -0.005880(71) & 884.47154(834) & 69.7$\pm$2.2 \\
\hline
OH 1720   & G1& -0.00287(23)$\times$1/9  & 912.3229(156) & 36.5$\pm$2.3 \\
          & G2& -0.005880(71)$\times$1/9 & 912.6726(181) & 69.7$\pm$2.2 \\
          & G3& 0.000566(74) & 912.3887(260) & 30.7$\pm$5.3  \\
          & G4& 0.000739(68) & 912.8025(249) & 39.0$\pm$4.5 \\

\hline
 \end{tabular}

\tablefoot{ From left to right are the line name, the name of OH gas, the normalized amplitude of Gaussian functions, the center of Gaussian functions, and the velocity width (FWHM) of Gaussian functions.
}
 
 \end{table*}

\begin{figure*}
        \includegraphics[width=\textwidth]{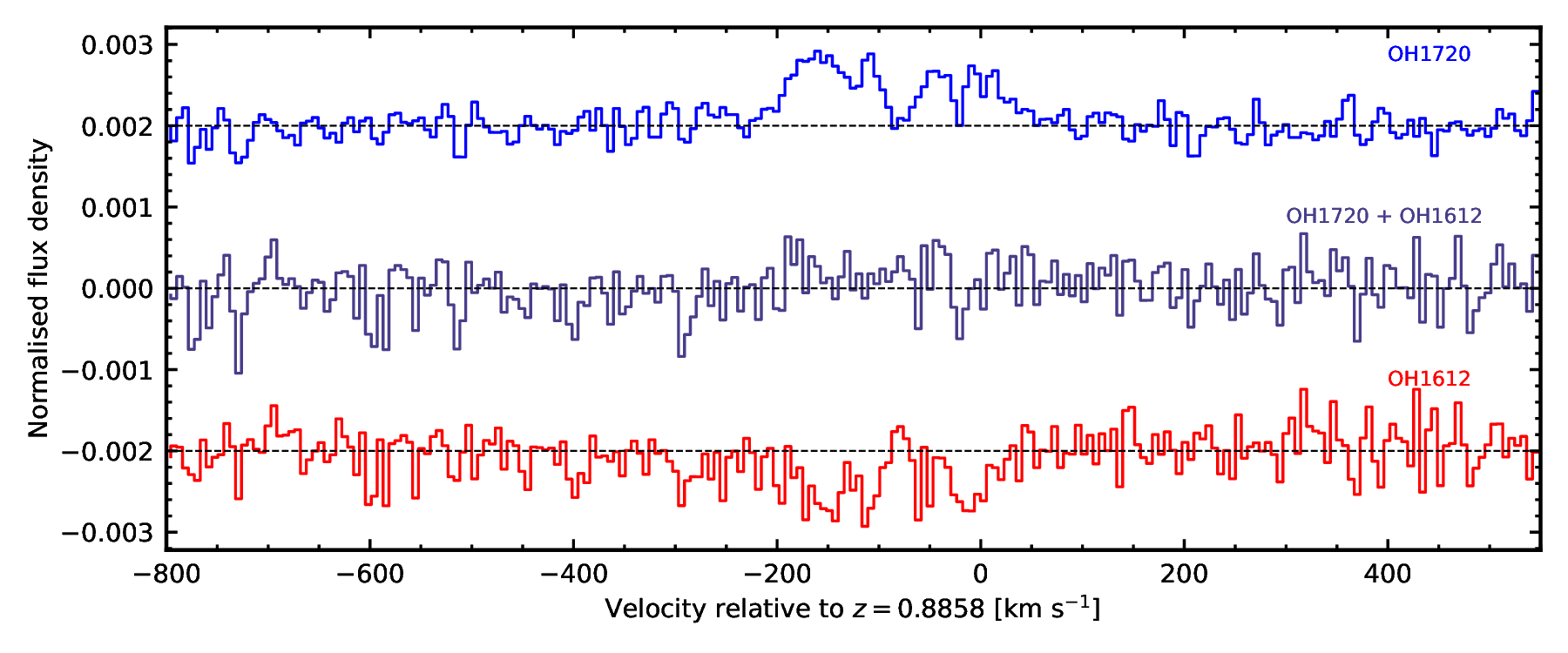}
    \caption{Conjugate OH satellite lines and their summed line. The top and bottom are OH satellite lines with the thermal counterparts subtracted, which have been shifted for clarity, while the middle is the summed spectrum. Both Kolmogorov-Smirnov
and Anderson-Darling tests have shown that the summed spectrum is consistent with being drawn from a Gaussian distribution, which strongly suggests that the OH satellite lines are conjugate.}
    \label{fig:OH1720+OH1612}
\end{figure*}

In the fitting, a crucial assumption is the existence of conjugate OH satellite lines. To test this, we subtracted the thermal absorption counterparts from the OH satellite lines. The remaining and summed spectra are presented in Fig. \ref{fig:OH1720+OH1612}. Both Kolmogorov-Smirnov and Anderson-Darling tests have shown that the summed spectrum is consistent with being drawn from a Gaussian distribution with p values of 0.64 and 0.46, which is strongly suggestive of a conjugate nature. We note that the feature in the residuals at 912.4 MHz has $\sim$ 3.7$\sigma$ significance, which is consistent with a noise peak.

\section{Measuring the cosmological evolution of fundamental constants} \label{sec:mc}

We measure the cosmological evolution of the fundamental constants $\alpha$, $\mu$, and $g_p$ using the measured redshifts of the OH 18 cm lines and the equations (Eq. 12 and 13) given in \cite{chengalur2003}. We note that the electron-proton mass ratio was used in \cite{chengalur2003} whereas the proton-electron mass ratio was used in later works, such as \cite{bagdonaite2013} and \cite{kanekar2018}. In this paper, we also define $\mu$ as the proton-electron mass ratio.

From the line fitting, we know that there are four OH gas clouds, of which two are thermalized while another two exhibit conjugate behavior. There are named G1, G2, G3, and G4 and their parameters are listed in Table \ref{tab:fit_results}. With the Eqs. (12) and (13) in \cite{chengalur2003}, comparisons of the redshifts between $\nu_{1667}$+$\nu_{1665}$ and $\nu_{1667}$-$\nu_{1665}$, and between $\nu_{1667}$+$\nu_{1665}$ and $\nu_{1720}$-$\nu_{1612}$ for G1 give

\begin{equation}\label{equ:G1}
 \rm G1\left\{
  \begin{array}{ll}
   0.13\frac{\Delta \mu}{\mu}+0.26\frac{\Delta \alpha}{\alpha}+\frac{\Delta g_p}{g_p}= (1.0\pm10.2)\times10^{-3}&  \text{(a)}  \\
   1.85\frac{\Delta \mu}{\mu}+3.70\frac{\Delta \alpha}{\alpha}+\frac{\Delta g_p}{g_p}=(-1.2\pm3.4)\times10^{-4}& \text{(b)}.
  \end{array}
  \right.
\end{equation}
For G2 we obtain
\begin{equation}\label{equ:G2}
 \rm G2\left\{
  \begin{array}{ll}
   0.13\frac{\Delta \mu}{\mu}+0.26\frac{\Delta \alpha}{\alpha}+\frac{\Delta g_p}{g_p}= (1.0\pm12.5)\times10^{-3}&  \text{(a)}  \\
   1.85\frac{\Delta \mu}{\mu}+3.70\frac{\Delta \alpha}{\alpha}+\frac{\Delta g_p}{g_p}=(-1.2\pm3.8)\times10^{-4}& \text{(b)}.
  \end{array}
  \right.
\end{equation}
For G3, a comparison of the redshifts between $\nu_{1720}$+$\nu_{1612}$ and $\nu_{1720}$-$\nu_{1612}$ gives
\begin{equation}\label{equ:G3}
1.85\frac{\Delta \mu}{\mu}+3.70\frac{\Delta \alpha}{\alpha}+\frac{\Delta g_p}{g_p}= (-1.2\pm5.3)\times10^{-4}
,\end{equation}
and for G4, we have
\begin{equation}\label{equ:G4}
1.85\frac{\Delta \mu}{\mu}+3.70\frac{\Delta \alpha}{\alpha}+\frac{\Delta g_p}{g_p}= (-1.2\pm5.0)\times10^{-4}
.\end{equation}
Combining Eq. \ref{equ:G1}(a) and \ref{equ:G2}(a), weighted based on their uncertainties, and combining Eqs. \ref{equ:G1}(b), \ref{equ:G2}(b), \ref{equ:G3}, and \ref{equ:G4}, weighted based on their uncertainties, give

\begin{equation}\label{equ:final}
 \left\{
  \begin{array}{ll}
   0.13\frac{\Delta \mu}{\mu}+0.26\frac{\Delta \alpha}{\alpha}+\frac{\Delta g_p}{g_p}= (1.0\pm7.9)\times10^{-3}&  \text{(a)}  \\
   1.85\frac{\Delta \mu}{\mu}+3.70\frac{\Delta \alpha}{\alpha}+\frac{\Delta g_p}{g_p}=(-1.2\pm2.1)\times10^{-4}& \text{(b)}.
  \end{array}
  \right.
\end{equation}
A comparison of various $\rm CH_{3}OH$ transitions in this absorber has given
$\Delta\mu/\mu=(-1.8\pm1.2)\times10^{-7}$ \citep{muller2021}.
Applying this to  Eq.~\ref{equ:final} gives 
\[
\frac{\Delta \alpha}{\alpha}+0.27\frac{g_p}{g_p}=\frac{\Delta (\alpha g_p^{0.27})}{(\alpha g_p^{0.27})}=(-3.2\pm5.7\times10^{-5}),
\]

\[
\frac{\Delta \alpha}{\alpha}=(-3.3\pm23.0)\times10^{-4}, 
\]

\[
\frac{\Delta g_p}{g_p}=(1.1\pm7.9)\times10^{-3}.
\]

\section{Discussion} \label{sec:discussion}

While measuring the fractional change in fundamental constants has been attempted with spectral lines from different species \citep[e.g.,][]{carilli2000,kanekar2011,kanekar2012}, velocity offsets between different species introduce redshift differences, which mimic a change in the fundamental constants. Using the lines from the same gas, such as the conjugate OH satellite lines \citep{darling2004,bagdonaite2013,kanekar2018}, avoids this systematic uncertainty. One example is in the  $z=0.8858$ absorption system toward PKS\,1830--211, where different transitions of the methanol molecule have provided $\Delta\mu/\mu=(-1.8\pm1.2)\times10^{-7}$ \citep{muller2021}. Here we use only the OH 18 cm lines from thermalized gas and the conjugate OH 18 cm satellite lines. 

Other potential sources of error could arise from the change in the continuum morphology of the background quasar \citep[e.g.,][]{garrett1997,lovell1998,martvidal2013} and variability in the absorption spectrum in the millimeter and sub-millimeter bands \citep[e.g.,][]{muller2008,muller2014,muller2023}. From 24 years of observations by various telescopes \citep{chengalur1999,koopmans2005,allison2017,gupta2021}, this variability does not appear to affect the centimeter band spectra; this result was also seen between the two MeerKAT observing runs \citep{gupta2021,combes2021}.  We therefore do not expect significant errors due to flux or spectral variability.

Compared to the constraint on $\mu$, the constraints on the evolution of $\alpha g_p^{0.27}$, $\alpha,$ and $g_p$ have a low precision. This is because the uncertainties are dominated by the 
$\nu_{1720}-\nu_{1612}$ and $\nu_{1667}-\nu_{1665}$ terms, which have redshift uncertainties one  and three orders of magnitude higher, respectively, than $\nu_{1667}+\nu_{1665}$ \citep{kanekar2004}. Using other lines, such as the OH main lines at other frequencies \citep{kanekar2004}, could increase the precision. However, there is as yet no detection of other OH lines toward PKS\,1830--211. \mbox{The H\,{\sc i}} absorption line was simultaneously detected with the OH 18 cm lines \citep{combes2021}. However, fitting the \mbox{H\,{\sc i}} line requires too many Gaussian functions, which makes it difficult to determine whether they truly represent individual components. Therefore, we excluded the \mbox{H\,{\sc i}} line in our analysis. Other molecular absorption lines were detected at frequencies much higher than OH 18 cm lines and show different profiles from OH 18 cm lines \citep[e.g.,][]{bottinelli2009,muller2014,muller2024}, due to the frequency-dependent radio structure of PKS 1830-211. Therefore, a comparison between OH 18 cm lines and other molecular lines is not appropriate. 

To date, measurements of the cosmological evolution of $g_p$ remain scarce. \cite{chengalur2003} reported the first simultaneous constraints on the variation of all the three fundamental constants by comparing line redshifts from different species, with uncertainties on the order of $\sim\,10^{-3}$. Since then, to our knowledge, no additional measurements of $g_p$ have been reported. In this work, we provide another simultaneous measurement of the cosmological evolution of all three fundamental constants based, however, on the same species.

\section{Summary} \label{sec:summary}

By utilizing the OH 18 cm absorption lines toward PKS\,1830--211 and 
combining these with previous measurements of the electron-proton mass ratio \citep{muller2021}, we 
have obtained (1$\sigma$) limits of 
\[
\frac{\Delta (\alpha g_p^{0.27})}{(\alpha g_p^{0.27})}\lesssim5.7\times10^{-5} \Rightarrow \lesssim7.8\times10^{-15}\,\rm yr^{-1},
\]

\[
\frac{\Delta \alpha}{\alpha}\lesssim2.3\times10^{-3} \Rightarrow \lesssim3.2\times10^{-13}\,\rm yr^{-1},
\]

\[
\Delta\mu/\mu\lesssim1.0\times10^{-7} \Rightarrow \lesssim1.4\times10^{-18}\,\rm yr^{-1} \text{ and}
\]

\[
\frac{\Delta g_p}{g_p}\lesssim7.9\times10^{-3}\Rightarrow \lesssim1.1\times10^{-12}\,\rm yr^{-1}
\]
over a look-back time of 7.3 Gyr.

\begin{acknowledgements}
The authors thank the anonymous referee for the comments that
improve this paper. The MeerKAT telescope is operated by the South African Radio Astronomy Observatory, which is a facility of the National Research Foundation, an agency of the Department of Science and Innovation. RSU acknowledges the support from National Science Foundation of China (grant 11988101) and the China Postdoctoral Science Foundation (Grant No. 2024M752979). DL is a New Cornerstone Investigator and is supported by the National Science Foundation of China (grant 11988101). MFG is supported by the National Science Foundation of China (grant 12473019), the China Manned Space Project with No. CMSCSST-2021-A06, the National SKA Program of China (Grant No. 2022SKA0120102), and the Shanghai Pilot Program for Basic Research-Chinese Academy of Science, Shanghai Branch (JCYJ-SHFY-2021-013).

\end{acknowledgements}

% WARNING
%-------------------------------------------------------------------
% Please note that we have included the references to the file aa.dem in
% order to compile it, but we ask you to:
%
% - use BibTeX with the regular commands:
%   \bibliographystyle{aa} % style aa.bst
%   \bibliography{Yourfile} % your references Yourfile.bib
%
% - join the .bib files when you upload your source files
%-------------------------------------------------------------------
\bibliographystyle{aa} % style aa.bst
\bibliography{aa53407-24corr} % your references Yourfile.bib

\end{document}